\documentclass[letterpaper, 10 pt, conference]{ieeeconf}
\IEEEoverridecommandlockouts
\usepackage{amsmath,amssymb,mathtools,bm}

\usepackage{amsthm}

\usepackage{algorithm}
\usepackage{algpseudocode}
\usepackage{cite}
\usepackage{url}
\usepackage{csquotes}

\usepackage{pgfplots}
\usepgfplotslibrary{groupplots}
\pgfplotsset{compat=1.18}

\newtheorem{theorem}{Theorem}
\newtheorem{lemma}{Lemma}
\newtheorem{proposition}{Proposition}
\newtheorem{corollary}{Corollary}
\theoremstyle{definition}
\newtheorem{remark}{Remark}

\DeclareMathOperator{\clip}{clip}
\DeclareMathOperator*{\argmin}{arg\,min}
\newcommand{\R}{\mathbb R}
\newcommand{\U}{\mathcal U}
\newcommand{\C}{\mathcal C}
\newcommand{\un}{u_{\mathrm{nom}}}
\newcommand{\ul}{u_{\min}}
\newcommand{\uu}{u_{\max}}
\newcommand{\ph}{\varphi}
\title{Finite-Step Algorithms for Input-Constrained Control Barrier Function-Based Safety Filters}
\author{Johannes Autenrieb$^{1}$, Mark Spiller$^{1}$, and Mayank S. Jha$^{2}$
\thanks{$^{1}$German Aerospace Center (DLR), Institute of Flight Systems, 38108 Braunschweig, Germany.
Email: \texttt{johannes.autenrieb@dlr.de, mark.spiller@dlr.de}.}
\thanks{$^{2}$CRAN, CNRS, Universit\'e de Lorraine, France.
Email: \texttt{mayank-shekhar.jha@univ-lorraine.fr}.}
}

\begin{document}
\maketitle
\begin{abstract}
Control barrier function (CBF) safety filters commonly rely on quadratic programs (QPs) to modify nominal control inputs while satisfying safety constraints. For safety-critical applications, however, the solution must be computed reliably under actuator limitations and with predictable computational effort. General-purpose QP solvers typically rely on numerical convergence criteria, which do not directly provide finite bounds on algorithmic steps or distinguish solver failure from mathematical infeasibility. This paper considers CBF-QPs with a single affine safety constraint, componentwise input bounds, and diagonal quadratic weights. Exploiting the established scalar dual structure of continuous quadratic knapsack problems, we obtain a necessary and sufficient pointwise feasibility condition and an exact breakpoint-based solution procedure. The algorithm requires at most $2n$ positive breakpoint events for $n$ inputs, reducing to at most $n$ events when the nominal input is admissible. In the latter case, an exact saturation--redistribution algorithm computes the optimal solution in at most $n$ passes. We further describe how the structured solution can be combined with verified floating-point constraint evaluation to distinguish certified feasible commands, certified infeasibility, and unresolved numerical cases. The resulting framework provides an exact and computationally predictable solution procedure for a practically relevant class of input-constrained CBF safety filters.  

\end{abstract}
\section{Introduction}
\label{sec:introduction}

Safety-critical control systems must enforce operating limits while preserving nominal control performance. Control barrier function (CBF) safety filters address this problem by modifying nominal inputs to satisfy safety constraints. In practical applications, however, actuator limitations can restrict the available control authority. For example, CBF-based flight-envelope protection may require corrective control actions while individual surfaces approach their deflection or rate limits \cite{Autenrieb_2025c,Hafner_2026}. The remaining actuators must then provide the required correction, if feasible. Distinguishing insufficient control authority from numerical solver failure is therefore essential for reliable safety-filter implementation.

CBF based safety filters are commonly formulated as quadratic programs (QPs) \cite{Ames2014ACC,Ames2017,TanDimarogonas2024Undesired}. Such safety filters have been developed recently, in an extensive manner, in various framework including safe reinforcement learning \cite{jha2026input,kanso2024off}.     Although input constraints can be incorporated directly, their active sets may change during optimization, complicating predictable real-time execution. General-purpose QP solvers typically terminate according to numerical convergence criteria, making it difficult to determine a priori the required computational effort or to distinguish numerical failure from an infeasible safety constraint. Recent studies have investigated explicit and resource-aware CBF filters \cite{Mestres2025ExplicitCBF}, structural feasibility and explicit solutions for particular system classes \cite{Mousavi2026Structure}, and conditions under which optimization-free projections recover the exact safety-filter solution \cite{Goel2026Exactness}. While these approaches exploit particular problem structures, they do not directly address the combined requirements of finite-step optimal control allocation, explicit feasibility assessment, and verified numerical constraint satisfaction for the input-constrained CBF-QPs considered here.

This paper considers CBF-QPs with a single affine safety constraint, componentwise actuator bounds, and diagonal positive-definite weights. The resulting problem belongs to the class of continuous quadratic knapsack problems, for which breakpoint-search and variable-fixing methods are well established \cite{Brucker1984,Kiwiel2008,Kiwiel2008VariableFixing}. Exploiting this structure for safety filtering provides a direct feasibility characterization and a finite-step computation of the minimally invasive control input under actuator constraints. For admissible nominal inputs, the solution also admits a saturation--redistribution interpretation, in which actuators reaching their relevant bounds are fixed and the remaining correction is redistributed among unsaturated actuators. Unlike general multi-axis control allocation \cite{Durham1993Constrained,Bodson2002,StephanFichter2019FastExact,HafnerSteinertHolzapfel2025Connecting}, where changes in the active set may require releasing previously saturated actuators, the single CBF constraint induces a monotone solution path without actuator release.

The contributions are threefold: (i) a necessary and sufficient pointwise feasibility condition and an exact breakpoint-based solution with at most $2n$ positive events; (ii) an exact saturation--redistribution algorithm requiring at most $n$ passes for admissible nominal inputs, together with explicit computational complexity bounds; and (iii) a finite-precision acceptance procedure distinguishing verified feasible commands, certified infeasibility, and unresolved numerical cases. Building on established continuous quadratic knapsack methods, the contribution lies in their structured application to input-constrained CBF safety filtering, including actuator position and rate limitations and numerical feasibility assessment. The resulting guarantees concern pointwise feasibility and bounded algorithmic work, not hardware worst-case execution times or automatic closed-loop safety guarantees \cite{RTCA2011DO178C}.

\section{Problem Formulation}
Consider the control-affine system
\begin{equation}\label{eq:system}
\dot x=f(x)+g(x)u,\qquad x\in\mathcal D\subseteq\R^{n_x},\quad u\in\R^n,
\end{equation}
where $f$ and $g$ are locally Lipschitz. Let $h:\mathcal D\to\R$ be continuously differentiable and let $\C=\{x:h(x)\geq0\}$. For an extended class-$\mathcal K$ function $\alpha$, the CBF condition is
\begin{equation}\label{eq:cbf}
L_fh(x)+L_gh(x)u\geq-\alpha(h(x)).
\end{equation}
At a fixed state define $a=(L_gh(x))^\top\in\R^n$ and $b=-L_fh(x)-\alpha(h(x))$. Thus \eqref{eq:cbf} is $a^\top u\geq b$.

The admissible control set is the nonempty box
\begin{equation}\label{eq:box}
\U=\{u\in\R^n:u_{\min,i}\leq u_i\leq u_{\max,i},\ i=1,\ldots,n\},
\end{equation}
with $u_{\min,i}<u_{\max,i}$. Given $\un\in\R^n$ and $W=\mathrm{diag}(w_1,\ldots,w_n)$, $w_i>0$, the filter solves
\begin{equation}\label{eq:qp}
\begin{aligned}
\min_{u\in\R^n}\quad &J(u)=\frac12(u-\un)^\top W(u-\un)\\
\mathrm{s.t.}\quad &a^\top u\geq b,\qquad u\in\U.
\end{aligned}
\end{equation}
Note that no assumption on $\un\in\U$ is made unless explicitly stated. Since $W\succ0$, \eqref{eq:qp} has a unique minimizer whenever feasible.

\subsection{Exact Feasibility Test}
For $a_i\ne0$, define the bound aligned with increasing $a^\top u$ by
\begin{equation}\label{eq:align}
\bar u_i=\begin{cases}u_{\max,i},&a_i>0,\\u_{\min,i},&a_i<0.\end{cases}
\end{equation}
For $a_i=0$ choose $\bar u_i=\clip(u_{\mathrm{nom},i},u_{\min,i},u_{\max,i})$, where $\clip(z,l,r)=\min\{r,\max\{l,z\}\}$.
\begin{proposition}[Pointwise feasibility]\label{prop:feas}
Problem \eqref{eq:qp} is feasible if and only if
\begin{equation}\label{eq:feas}
b\leq M:=a^\top\bar u=\sum_{i=1}^n\max\{a_iu_{\min,i},a_iu_{\max,i}\}.
\end{equation}
\end{proposition}
\begin{proof}
Maximization of $a^\top u$ over \eqref{eq:box} separates across the coordinates. Each term is maximized at its aligned bound, so $M=\max_{u\in\U}a^\top u$. Therefore a point in $\U$ satisfies $a^\top u\geq b$ exactly when $M\geq b$.
\end{proof}
If $a=0$, the condition becomes $b\leq0$; if feasible, the solution is simply $\clip(\un,\ul,\uu)$ componentwise. A feasible pointwise CBF inequality by itself does not establish that the closed-loop system can satisfy it for all future times.

\subsection{Actuator Position and Rate Constraints}

The formulation also accommodates actuator-rate constraints by considering the control increment $v=u-u_{\mathrm{prev}}$, where $u_{\mathrm{prev}}$ denotes the previously applied input. For a sampling period $T_s>0$, componentwise position and rate limits define the admissible increment bounds
\begin{equation}
\begin{aligned}
v_{\min,i}
&=\max\left\{
u_{\min,i}-u_{\mathrm{prev},i},
T_s\dot u_{\min,i}
\right\},\\
v_{\max,i}
&=\min\left\{
u_{\max,i}-u_{\mathrm{prev},i},
T_s\dot u_{\max,i}
\right\}.
\end{aligned}
\label{eq:increment_bounds}
\end{equation}
Assuming the resulting increment box is nonempty, the CBF inequality becomes
\begin{equation}
a^\top v\geq b-a^\top u_{\mathrm{prev}}.
\label{eq:increment_cbf}
\end{equation}
Setting $v_{\mathrm{nom}}=u_{\mathrm{nom}}-u_{\mathrm{prev}}$, the quadratic objective satisfies
\begin{equation}
\frac12(v-v_{\mathrm{nom}})^\top W(v-v_{\mathrm{nom}})
=\frac12(u-u_{\mathrm{nom}})^\top W(u-u_{\mathrm{nom}}).
\end{equation}
Thus, the incremental formulation preserves the original objective. The proposed algorithms therefore apply with nominal input $v_{\mathrm{nom}}$, threshold $b-a^\top u_{\mathrm{prev}}$, and the increment bounds in \eqref{eq:increment_bounds}.

\section{Scalar Dual Path and Breakpoints}
The key simplification, also familiar from continuous quadratic knapsack optimisation \cite{Kiwiel2008}, is to replace a search over $n$ control components by a search over one scalar. Introduce a nonnegative multiplier $\lambda$ for the CBF inequality and define
\begin{equation}\label{eq:path}
\begin{aligned}
u(\lambda)&=\argmin_{u\in\U}\{J(u)-\lambda a^\top u\},\\
\ph(\lambda)&=a^\top u(\lambda).
\end{aligned}
\end{equation}
For a fixed $\lambda$, each input can be computed independently: first shift the nominal command by $\lambda a_i/w_i$, then clip the result to its limits. Because the weights and bounds are separable,
\begin{equation}\label{eq:clip}
u_i(\lambda)=\clip\left(u_{\mathrm{nom},i}+\lambda\frac{a_i}{w_i},u_{\min,i},u_{\max,i}\right).
\end{equation}
The following lemma recalls the monotonicity and piecewise-affine properties of the dual path underlying continuous quadratic knapsack algorithms \cite{Kiwiel2008}.
\begin{lemma}[Monotone dual path]\label{lem:mono}
$\ph$ is continuous, piecewise affine, and nondecreasing on $[0,\infty)$. The monotonicity statement remains valid for any positive definite $W$ and any closed convex $\U$ for which the minimizers in \eqref{eq:path} exist.
\end{lemma}
\begin{proof}
For $\lambda_2>\lambda_1\geq0$, optimality of $u_k=u(\lambda_k)$ gives $J(u_1)-\lambda_1a^\top u_1\leq J(u_2)-\lambda_1a^\top u_2$ and $J(u_2)-\lambda_2a^\top u_2\leq J(u_1)-\lambda_2a^\top u_1$. Adding yields $(\lambda_2-\lambda_1)(\ph(\lambda_2)-\ph(\lambda_1))\geq0$. Continuity and piecewise affinity follow from \eqref{eq:clip}.
\end{proof}

For $a_i\ne0$, let $\check u_i$ denote the bound opposite to $\bar u_i$ and set
\begin{equation}\label{eq:break}
\lambda_i^{\rm rel}=\frac{w_i(\check u_i-u_{\mathrm{nom},i})}{a_i},\qquad
\lambda_i^{\rm sat}=\frac{w_i(\bar u_i-u_{\mathrm{nom},i})}{a_i}.
\end{equation}
In both sign cases, $\lambda_i^{\rm rel}<\lambda_i^{\rm sat}$. Along the path, coordinate $i$ is at the opposed bound below $\lambda_i^{\rm rel}$, is free between these breakpoints, and is at its aligned bound above $\lambda_i^{\rm sat}$. Events at nonpositive multipliers are absorbed into the initial state. The right derivative is
\begin{equation}\label{eq:slope}
\ph'_+(\lambda)=\sum_{\substack{i:a_i\ne0,\\\lambda_i^{\rm rel}\leq\lambda<\lambda_i^{\rm sat}}}\frac{a_i^2}{w_i}.
\end{equation}
The right-limit convention in \eqref{eq:slope} matters when $u_{\mathrm{nom},i}$ lies exactly at an opposed bound, so $\lambda_i^{\rm rel}=0$: that coordinate becomes free immediately for $\lambda>0$.

\begin{lemma}[Breakpoint count]\label{lem:break}
There are at most $2n$ strictly positive release or saturation events on the path. If $\un\in\U$, every $\lambda_i^{\rm rel}\leq0$, so there are at most $n$ strictly positive events, all of which remove coordinates from the free set.
\end{lemma}
\begin{proof}
Equation \eqref{eq:clip} is the clipping of an affine function whose direction is fixed by $a_i$. Hence a coordinate can pass from the opposed bound to the interior and from the interior to the aligned bound at most once each. If $u_{\mathrm{nom},i}\in[u_{\min,i},u_{\max,i}]$, its starting value cannot lie beyond the opposed bound, which makes $\lambda_i^{\rm rel}\leq0$.
\end{proof}

\begin{remark}
Simply replacing an inadmissible $\un$ by $\clip(\un,\ul,\uu)$ is not generally equivalent to \eqref{eq:qp}: it changes the objective and can change the optimal distribution of CBF correction across actuators. The $2n$-event case is therefore useful even if upstream nominal-command limiting is available.
\end{remark}

\section{Exact Solution by a Breakpoint Sweep}
\label{sec:exact_solution}\textit{}

The scalar dual path provides a direct way to solve the CBF-QP without iterative convergence tests. Let $u(0)=\clip(\un,\ul,\uu)$ and $\ph_0=a^\top u(0)$. If $\ph_0\geq b$, the box-constrained nominal input already satisfies the CBF inequality and is therefore optimal. Otherwise, the smallest multiplier $\lambda>0$ satisfying $\ph(\lambda)=b$ determines the optimal correction, except at the terminal feasible boundary $b=M$, where the optimal input is the aligned maximizer $\bar u$.

Since $\ph$ is continuous and piecewise affine, the multiplier can be determined by examining its breakpoints. A release event at $\lambda_i^{\rm rel}>0$ increases the slope of $\ph$ by $a_i^2/w_i$, whereas a saturation event at $\lambda_i^{\rm sat}>0$ decreases it by the same amount. Events occurring at the same multiplier are processed together, while events at zero are already accounted for by the initial right derivative in \eqref{eq:slope}. This leads to Algorithm~\ref{alg:sweep}.

\begin{algorithm}[t]
\caption{Exact-arithmetic breakpoint filter}
\label{alg:sweep}
\begin{algorithmic}[1]
\Require $a,b,\un,\ul,\uu,w$ with $w_i>0$
\State $u_0\gets\clip(\un,\ul,\uu)$
\State $\ph\gets a^\top u_0$
\If{$\ph\geq b$}
    \State \Return $u_0$, feasible
\EndIf
\State Form $\bar u$ using \eqref{eq:align}
\State $M\gets a^\top\bar u$
\If{$M<b$}
    \State \Return infeasible
\EndIf
\State Form and sort all positive events from \eqref{eq:break}
\State $\lambda\gets0$; $s\gets\ph'_+(0)$
\For{each distinct event value $t$ in increasing order}
    \If{$s>0$ \textbf{and} $b\leq\ph+s(t-\lambda)$}
        \State $\lambda^\star\gets\lambda+(b-\ph)/s$
        \State \Return $u(\lambda^\star)$, feasible
    \EndIf
    \State $\ph\gets\ph+s(t-\lambda)$
    \State $\lambda\gets t$
    \State Update $s$ using all events at $t$
\EndFor
\State \Return $\bar u$, feasible
\Comment{$b=M$ at terminal plateau}
\end{algorithmic}
\end{algorithm}

\begin{theorem}[Exact solution and finite event bound]
\label{thm:main}
Suppose $W$ is diagonal and positive definite and the admissible input set is the nonempty box \eqref{eq:box}. In exact arithmetic, Algorithm~\ref{alg:sweep} returns the unique minimizer of \eqref{eq:qp} whenever \eqref{eq:feas} holds and otherwise correctly identifies infeasibility. The algorithm processes at most $2n$ strictly positive breakpoint events. If $\un\in\U$, this bound reduces to $n$. The result also holds at the terminal feasible boundary $b=M$.
\end{theorem}

\begin{proof}
Feasibility is characterized by Proposition~\ref{prop:feas}. If $\ph_0\geq b$, then $u(0)$ minimizes $J$ over $\U$ and is therefore also optimal for \eqref{eq:qp}.

Otherwise, $\ph_0<b\leq M$. Between consecutive breakpoints, $\ph$ is affine with a nonnegative slope $s$ given by \eqref{eq:slope}. Algorithm~\ref{alg:sweep} either finds the first intersection with $b$ on the current interval or advances to the next breakpoint. At an intersection $\lambda^\star$, the corresponding input $u^\star=u(\lambda^\star)$ satisfies $a^\top u^\star=b$ and minimizes $J(u)-\lambda^\star a^\top u$ over $\U$. Hence,
\begin{equation}
J(v)-\lambda^\star a^\top v
\geq
J(u^\star)-\lambda^\star a^\top u^\star
\label{eq:optineq}
\end{equation}
for every $v\in\U$. Since $\lambda^\star\geq0$ and $a^\top v\geq b=a^\top u^\star$ for every feasible $v$, it follows that $J(v)\geq J(u^\star)$. Strict convexity establishes uniqueness.

If no interval reaches $b$, the final value of $\ph$ is $M$, so feasibility requires $b=M$. Every feasible input must then maximize $a^\top u$ over $\U$. Thus, $u_i=\bar u_i$ for $a_i\ne0$, while the coordinates with $a_i=0$ minimize their individual quadratic terms and are given by $\clip(u_{\mathrm{nom},i},u_{\min,i},u_{\max,i})$. Consequently, $\bar u$ is again the unique minimizer. The event bounds follow from Lemma~\ref{lem:break}.
\end{proof}

\subsection{Saturation and Redistribution}

When $\un\in\U$, the dual path starts from an admissible nominal command. As $\lambda$ increases, each actuator moves toward the bound that maximizes its contribution to the CBF inequality. Once an actuator reaches this aligned bound, it remains saturated for all larger multipliers.

This property permits a simple redistribution procedure. Let $F$ denote the set of unsaturated actuators. While the active set remains unchanged, the increase in the barrier derivative is governed by
\begin{equation}
S_F=\sum_{i\in F}\frac{a_i^2}{w_i}.
\label{eq:gain}
\end{equation}
For a remaining CBF deficit $d=b-a^\top u$, the required unconstrained correction is
\begin{equation}
\Delta\lambda=\frac{d}{S_F},
\qquad
\Delta u_i=\frac{a_i}{w_i}\Delta\lambda,
\quad i\in F.
\label{eq:redistribution_step}
\end{equation}
If this correction exceeds an actuator bound, the corresponding input is fixed at its aligned limit and the remaining correction is redistributed among the other actuators.

This resembles saturation--redistribution methods used in control allocation \cite{Durham1993Constrained,Bodson2002,StephanFichter2019FastExact,HafnerSteinertHolzapfel2025Connecting}. The important simplification is that the single CBF constraint induces a monotone scalar dual path. Consequently, an actuator that reaches its aligned bound does not need to be released during subsequent redistribution steps.

\begin{algorithm}[t]
\caption{Saturation--redistribution ($\un\in\U$)}
\label{alg:redist}
\begin{algorithmic}[1]
\Require $a,b,\un,\ul,\uu,w$ with $w_i>0$, $\un\in\U$
\If{$a^\top\un\geq b$}
    \State \Return $\un$, feasible
\EndIf
\State Form $\bar u$ using \eqref{eq:align}
\If{$a^\top\bar u<b$}
    \State \Return infeasible
\EndIf
\State $F\gets\{i:a_i\ne0\}$; $S\gets\emptyset$
\For{$k=1,\ldots,n$}
    \State $D\gets\sum_{i\in F}a_i^2/w_i$
    \If{$D=0$}
        \State \Return $\bar u$, feasible
    \EndIf
    \State $\lambda\gets
    \bigl(b-\sum_{i\in S}a_i\bar u_i
    -\sum_{i\in F}a_i u_{\mathrm{nom},i}\bigr)/D$
    \State $\tilde u_i\gets u_{\mathrm{nom},i}+\lambda a_i/w_i$, $i\in F$
    \State $V\gets\{i\in F:a_i(\tilde u_i-\bar u_i)>0\}$
    \If{$V=\emptyset$}
        \State $u_i\gets\tilde u_i$, $i\in F$
        \State $u_i\gets\bar u_i$, $i\in S$
        \State $u_i\gets u_{\mathrm{nom},i}$, $a_i=0$
        \State \Return $u$, feasible
    \EndIf
    \State $S\gets S\cup V$
    \State $F\gets F\setminus V$
\EndFor
\State \Return $\bar u$, feasible
\Comment{Terminal feasible boundary}
\end{algorithmic}
\end{algorithm}

\begin{proposition}[Exact redistribution and bounded passes]
\label{prop:redist}
Under the assumptions of Theorem~\ref{thm:main}, with $\un\in\U$ and $a^\top\bar u\geq b$, Algorithm~\ref{alg:redist} returns the unique minimizer of \eqref{eq:qp} in exact arithmetic. The algorithm requires at most $n$ redistribution passes.
\end{proposition}

\begin{proof}
If the nominal input satisfies the CBF inequality, it is optimal. Otherwise, the optimal multiplier is positive, and each coordinate with $a_i\ne0$ moves monotonically toward its aligned bound along \eqref{eq:clip}.

For a given free set $F$, Algorithm~\ref{alg:redist} computes the multiplier that would satisfy the CBF equality if the remaining actuators were unconstrained. If the resulting trial input respects all aligned bounds, it satisfies the KKT conditions of \eqref{eq:qp} and is therefore optimal.

Otherwise, let $V\subseteq F$ denote the violating coordinates. If $F\setminus V\neq\emptyset$, the multiplier after fixing these coordinates satisfies
\begin{equation}
\lambda_{\mathrm{new}}-\lambda
=
\frac{\displaystyle\sum_{i\in V}a_i(\tilde u_i-\bar u_i)}
{\displaystyle\sum_{i\in F\setminus V}a_i^2/w_i}>0.
\end{equation}
The numerator is positive by the definition of $V$. Hence, the multiplier increases, and previously saturated actuators remain at their aligned bounds. Since $\un\in\U$ and $\lambda\geq0$, no free actuator crosses its opposite bound. Fixing the violating coordinates therefore preserves the optimal solution while reducing the free set. Each unsuccessful pass removes at least one actuator, so the algorithm terminates within at most $n$ passes. If no free actuator remains, feasibility implies $b=M$, and the aligned maximizer is optimal. Coordinates with $a_i=0$ remain at their nominal values throughout. Uniqueness follows from strict convexity of the objective.
\end{proof}

\subsection{Computational Complexity and Predictability}

The finite-event bounds established above describe the number of changes along the scalar dual path, but they should be distinguished from the total computational effort.

Algorithm~\ref{alg:sweep} generates at most $2n$ positive breakpoints. Sorting these events requires $O(n\log n)$ comparisons using conventional comparison sorting, while the subsequent sweep requires $O(n)$ operations when coincident events are processed together. The overall complexity is therefore $O(n\log n)$.

For admissible nominal inputs, Algorithm~\ref{alg:redist} avoids sorting and instead repeatedly updates the remaining free set. Each unsuccessful pass removes at least one actuator, yielding at most $n$ passes. With direct recomputation of the required sums and saturation tests, its worst-case arithmetic complexity is $O(n^2)$. Both algorithms thus provide computational bounds determined by the number of control inputs rather than by convergence tolerances. For implementations requiring a fixed execution schedule, the breakpoint algorithm can additionally employ a predetermined sorting network \cite{Batcher1968Sorting} and fixed-count processing of a padded event list. Such an implementation bounds the number of comparisons and arithmetic operations, but does not by itself establish a hardware worst-case execution time. The latter requires analysis of the compiled implementation, including floating-point operations and exceptional-value handling.

\section{Verifiable Floating-Point Implementation}
\label{sec:verification}

The preceding algorithms provide exact solutions in real arithmetic. In a floating-point implementation, however, rounding errors affect the computation of breakpoints, multipliers, and control inputs. These errors become particularly relevant near actuator limits or the feasibility boundary, where small numerical deviations may change the sign of a constraint residual. Consequently, a finite bound on breakpoint events does not guarantee that the computed input satisfies the CBF inequality. In particular, observing
$\operatorname{fl}(a^\top\hat u-b)\geq0$
does not necessarily imply
$a^\top\hat u\geq b$
in exact arithmetic \cite{Higham2002Accuracy}.

To address this issue, candidate generation and constraint verification are treated separately. Algorithms~\ref{alg:sweep} and \ref{alg:redist} compute a candidate input $\hat u$ using floating-point arithmetic. The candidate is subsequently accepted only if satisfaction of the represented CBF inequality and actuator bounds can be established independently of the numerical decisions made during optimization.

\subsection{Verified Constraint Evaluation}

Consider the exact real values represented by the stored floating-point coefficients $a$, $b$, input bounds, and candidate $\hat u$. Define the CBF residual and actuator slacks as
\begin{equation}
\begin{aligned}
R(\hat u)&=a^\top\hat u-b,\\
d_i^-(\hat u)&=\hat u_i-u_{\min,i},\\
d_i^+(\hat u)&=u_{\max,i}-\hat u_i.
\end{aligned}
\label{eq:verification_residuals}
\end{equation}
Ordinary floating-point evaluation of these expressions may produce incorrect signs close to zero.

Instead, verified interval arithmetic is used to compute rigorous lower bounds
\begin{equation}
\begin{aligned}
\underline R(\hat u)&\leq R(\hat u),\\
\underline d_i^-(\hat u)&\leq d_i^-(\hat u),\\
\underline d_i^+(\hat u)&\leq d_i^+(\hat u).
\end{aligned}
\label{eq:interval}
\end{equation}
For example, an enclosure of $R(\hat u)$ can be obtained by evaluating the products $a_i\hat u_i$, their sum, and the subtraction of $b$ using outward rounding. The lower endpoint of the resulting interval provides $\underline R(\hat u)$. The box slacks are evaluated analogously. This procedure requires an arithmetic implementation that guarantees outward enclosures, including appropriate handling of exceptional floating-point values.

\begin{proposition}[Verified acceptance]
\label{prop:verify}
Suppose that the bounds in \eqref{eq:interval} rigorously enclose the exact residuals of the represented floating-point data. If a finite candidate $\hat u$ satisfies
\begin{equation}
\underline R(\hat u)\geq0,\qquad
\underline d_i^-(\hat u)\geq0,\qquad
\underline d_i^+(\hat u)\geq0
\label{eq:acceptance}
\end{equation}
for all $i=1,\ldots,n$, then $\hat u$ satisfies the represented CBF inequality and all actuator bounds in exact real arithmetic.
\end{proposition}

\begin{proof}
Each verified lower bound does not exceed the corresponding exact constraint residual. Nonnegativity of all lower bounds therefore implies nonnegativity of the exact CBF residual and actuator slacks.
\end{proof}

\subsection{Feasibility Decisions and Numerical Limitations}

The acceptance test is sufficient but not necessary for feasibility. In particular, a mathematically feasible command may remain unresolved if rounding prevents its residual from being certified as nonnegative. Furthermore, verified feasibility does not establish that the floating-point candidate is the exact minimizer of \eqref{eq:qp}.

If the candidate produced by Algorithm~\ref{alg:sweep} or \ref{alg:redist} cannot be verified, the aligned maximizer $\bar u$ from \eqref{eq:align} provides a possible fallback candidate. Since $\bar u$ maximizes $a^\top u$ over $\U$, it achieves the largest possible CBF residual, although it need not minimize the quadratic objective.

Infeasibility can be certified separately. Let $M=a^\top\bar u$ denote the maximal attainable constraint value. If a verified upper bound $\overline R_{\max}$ satisfies
\begin{equation}
M-b\leq\overline R_{\max}<0,
\label{eq:verified_infeasibility}
\end{equation}
then the represented CBF-QP is infeasible. If neither a feasible candidate nor infeasibility can be certified, the numerical outcome must be classified as unresolved. This situation can arise near the terminal feasibility boundary $b=M$, where the exact problem is feasible but numerical enclosures may contain zero.

The proposed verification therefore distinguishes three outcomes: verified feasible, verified infeasible, and unresolved. It prevents an incorrectly rounded residual from being accepted as evidence of feasibility, but does not guarantee that every feasible problem can be resolved in finite precision. Moreover, the certificate applies only to the represented problem data. Uncertainty in the dynamics, state estimation, actuator tracking, and intersample evolution requires additional analysis and is not covered by the numerical acceptance test.

\section{Implications for Closed-Loop Safety}

The preceding results establish pointwise optimality and feasibility. Closed-loop safety additionally requires that the applied input satisfies the CBF condition along a well-defined system trajectory.

\begin{corollary}[Conditional forward invariance]
Suppose a closed-loop solution of \eqref{eq:system} exists on $[0,T)$ with $x(0)\in\mathcal C$, and the applied control satisfies $u(t)\in\mathcal U$ and $a(x(t))^\top u(t)\geq b(x(t))$ almost everywhere. Under the standard comparison conditions on $\alpha$, the solution remains in $\mathcal C$ on $[0,T)$.
\end{corollary}

\begin{proof}
The CBF condition implies $\dot h(x(t))\geq-\alpha(h(x(t)))$ almost everywhere. Forward invariance follows from the comparison principle.
\end{proof}

Although the pointwise optimizer is unique, its regularity as a state-dependent feedback requires additional conditions. In particular, the dual multiplier may be ill-conditioned near the terminal feasible boundary $b=M$. Local Lipschitz continuity follows on fixed-active-set regions with locally Lipschitz problem data and a dual-path slope bounded away from zero, but is not established globally. Furthermore, samplewise constraint satisfaction alone does not guarantee intersample safety.

\section{Numerical Results}
\label{sec:results}

\subsection{Randomized QP Benchmark}

The computational performance of the proposed algorithms was evaluated using randomized instances of the CBF-QP in \eqref{eq:qp}. Since the algorithms depend only on the pointwise quantities $a$, $b$, $W$, and the actuator bounds, the benchmark considers these quantities directly, independently of a particular dynamic system. This allows the solution procedures to be evaluated over different input dimensions, active-set configurations, and feasibility conditions.

The number of inputs was varied over $n\in\{4,8,16,32\}$. For each problem, the constraint coefficients were drawn from $a_i\sim\mathcal N(0,1)$, while the diagonal weights and input bounds were generated independently according to
\begin{equation}
\begin{aligned}
w_i &\sim \mathcal U(0.2,3.0),\\
u_{\min,i} &\sim \mathcal U(-1.0,-0.3),\\
u_{\max,i} &\sim \mathcal U(0.3,1.0).
\end{aligned}
\label{eq:benchmark_sampling}
\end{equation}
Here, $\mathcal U$ denotes the uniform distribution. Both nominal inputs satisfying $u_{\mathrm{nom}}\in\mathcal U$ and nominal inputs outside the actuator bounds were considered.

To examine different feasibility conditions, the CBF threshold $b$ was selected relative to the maximum attainable constraint value $M$ from \eqref{eq:feas}. Four cases were considered: feasible problems away from the terminal boundary, nearly terminal feasible problems, terminal problems with $b=M$, and infeasible problems with $b>M$. The near-terminal and infeasible cases used offsets of $10^{-10}$ and $10^{-8}$, respectively, scaled by $\max\{1,|M|\}$. With $80$ realizations per dimension, nominal-input category, and feasibility case, the benchmark comprised $2{,}560$ problems, including $1{,}920$ feasible and $640$ infeasible instances.

The breakpoint sweep was evaluated for all problems, while the redistribution algorithm was applied only when $u_{\mathrm{nom}}\in\mathcal U$. For comparison, MATLAB's \texttt{quadprog} was used with its \texttt{interior-point-convex} and \texttt{active-set} algorithms. Both reference methods solved the same weighted QP, with the active-set algorithm initialized at the box-clipped nominal input.

\begin{figure*}[t]
\centering
\begin{tikzpicture}
\begin{groupplot}[
 group style={group size=1 by 3,vertical sep=1.15cm},
 width=0.82\textwidth,height=3.25cm,
 xmin=3,xmax=33,xtick={4,8,16,32},
 grid=major,grid style={black!12},
 tick label style={font=\footnotesize},label style={font=\footnotesize},
 legend style={font=\scriptsize,draw=black!30,fill=white,legend columns=4},
 every axis plot/.append style={thick,mark size=2.0pt}]
\nextgroupplot[ymin=0,ymax=1,ylabel={Time [ms]},title={(a) Median runtime},legend pos=north west]
\addplot+[blue,mark=o] table[x=n,y=sweep,col sep=comma]{data/benchmark.csv};\addlegendentry{Sweep}
\addplot+[teal!75!black,mark=square*] table[x=n,y=redis,col sep=comma]{data/benchmark.csv};\addlegendentry{Redistribution$^{\dagger}$}
\addplot+[orange!90!black,mark=triangle*] table[x=n,y=ip,col sep=comma]{data/benchmark.csv};\addlegendentry{Interior-point}
\addplot+[violet,mark=diamond*] table[x=n,y=active,col sep=comma]{data/benchmark.csv};\addlegendentry{Active-set}
\nextgroupplot[ymin=0,ymax=70,ylabel={Steps},title={(b) Maximum observed algorithmic steps},legend pos=north west]
\addplot+[blue,mark=o] table[x=n,y=maxSweep,col sep=comma]{data/benchmark.csv};\addlegendentry{Sweep events}
\addplot+[teal!75!black,mark=square*] table[x=n,y=maxRedis,col sep=comma]{data/benchmark.csv};\addlegendentry{Redistribution passes$^{\dagger}$}
\addplot+[black,dashed,mark=none] table[x=n,y=b2n,col sep=comma]{data/benchmark.csv};\addlegendentry{Sweep bound $2n$}
\addplot+[gray!75!black,densely dotted,mark=none] table[x=n,y=bn,col sep=comma]{data/benchmark.csv};\addlegendentry{Redistribution bound $n$}
\nextgroupplot[ymin=97.5,ymax=100.5,ylabel={Solved cases [\%]},xlabel={Number of inputs $n$},title={(c) Successfully solved feasible QPs},legend pos=south west]
\addplot+[blue,mark=o] table[x=n,y=sweep,col sep=comma]{data/outcomes.csv};\addlegendentry{Sweep}
\addplot+[teal!75!black,mark=square*] table[x=n,y=redis,col sep=comma]{data/outcomes.csv};\addlegendentry{Redistribution$^{\dagger}$}
\addplot+[orange!90!black,mark=triangle*] table[x=n,y=ip,col sep=comma]{data/outcomes.csv};\addlegendentry{Interior-point}
\addplot+[violet,mark=diamond*] table[x=n,y=active,col sep=comma]{data/outcomes.csv};\addlegendentry{Active-set}
\end{groupplot}
\end{tikzpicture}
\caption{Runtime, algorithmic steps, and feasible-case solution rates in the randomized benchmark. $^{\dagger}$Redistribution applies only when $u_{\mathrm{nom}}\in\mathcal U$.}
\label{fig:benchmark}
\end{figure*}
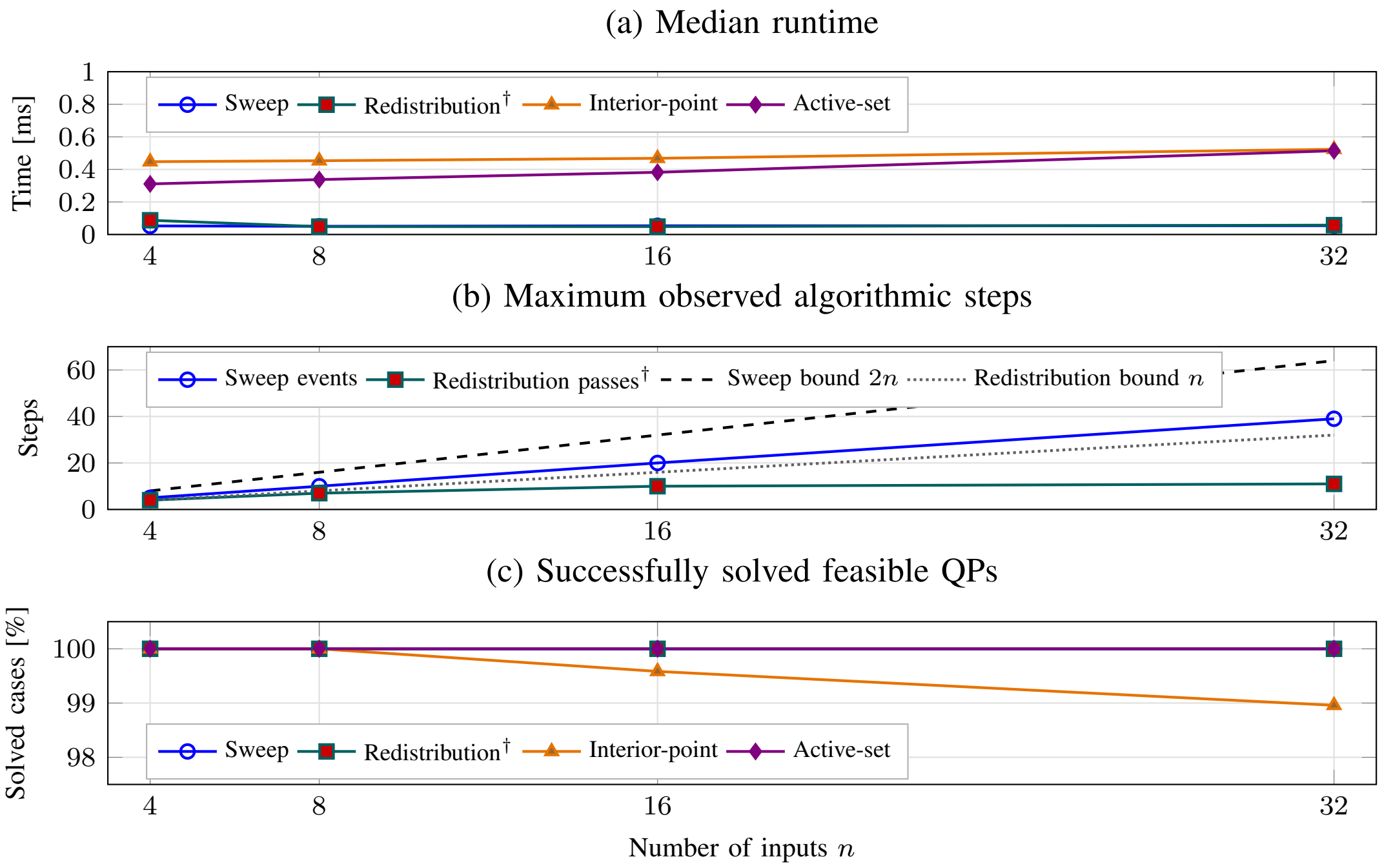

Figure~\ref{fig:benchmark}(a) compares the median observed execution times. The proposed algorithms achieve lower runtimes than both general-purpose QP methods over the considered dimensions. At $n=32$, the breakpoint sweep required approximately $0.054\,\mathrm{ms}$, compared with $0.524\,\mathrm{ms}$ and $0.514\,\mathrm{ms}$ for the interior-point and active-set methods, respectively. The redistribution algorithm achieved a median runtime of approximately $0.058\,\mathrm{ms}$ on its applicable subset.

Figure~\ref{fig:benchmark}(b) shows the maximum observed algorithmic work. For $n=4,8,16,32$, the sweep processed at most $5$, $10$, $20$, and $39$ positive breakpoint events, respectively, remaining below the theoretical bound of $2n$. The redistribution algorithm required at most $4$, $7$, $10$, and $11$ passes, respectively, consistent with the bound of $n$. These bounds concern different algorithmic operations and do not directly represent processor instruction counts.

Figure~\ref{fig:benchmark}(c) summarizes the solver success rates. The breakpoint sweep and active-set method successfully solved all $1{,}920$ feasible problems, while redistribution solved all eligible cases. The interior-point method failed to report successful termination in seven cases, all near the terminal feasibility boundary $b=M$, including five failures to compute a search direction. This indicates numerical sensitivity near the limit of attainable control authority rather than a general loss of accuracy with increasing dimension. The breakpoint sweep additionally identified all $640$ infeasible instances.

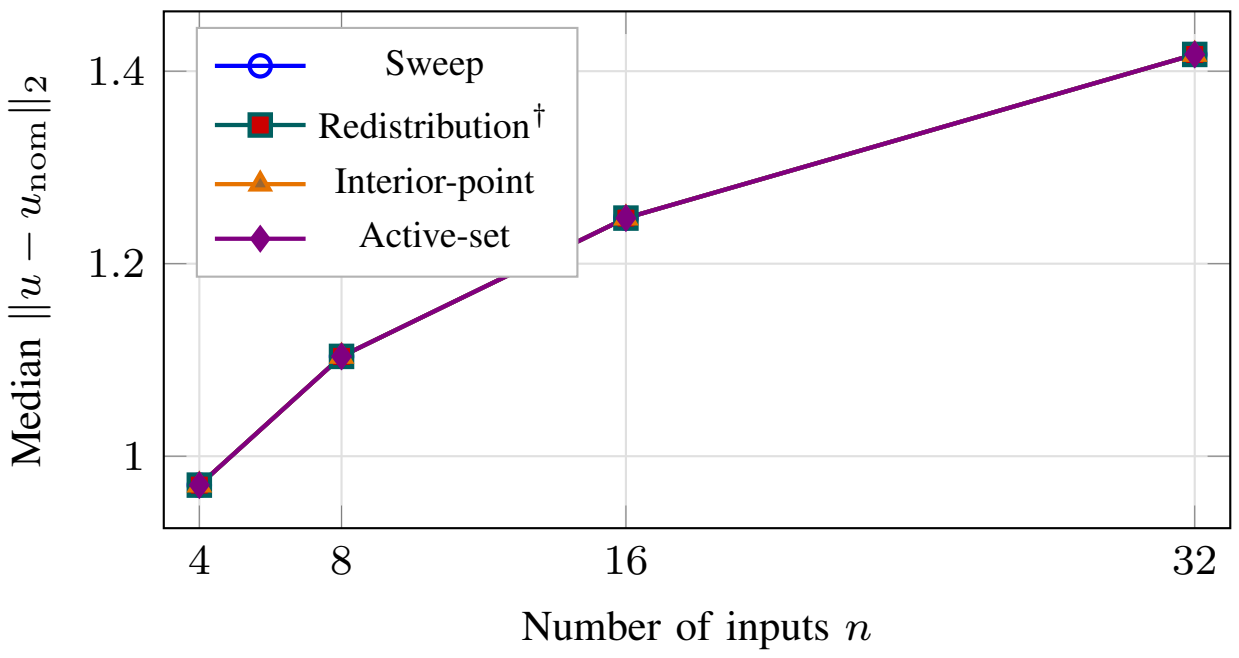
\begin{figure}[t]
\centering
\begin{tikzpicture}
\begin{axis}[width=\columnwidth,height=5.0cm,xmin=3,xmax=33,xtick={4,8,16,32},
 xlabel={Number of inputs $n$},ylabel={Median $\|u-u_{\mathrm{nom}}\|_2$},
 grid=major,grid style={black!12},tick label style={font=\footnotesize},
 label style={font=\footnotesize},legend style={font=\scriptsize,draw=black!30,fill=white},legend pos=north west]
\addplot+[blue,thick,mark=o] table[x=n,y=sweep,col sep=comma]{data/euclidean_corrections.csv};\addlegendentry{Sweep}
\addplot+[teal!75!black,thick,mark=square*] table[x=n,y=redis,col sep=comma]{data/euclidean_corrections.csv};\addlegendentry{Redistribution$^{\dagger}$}
\addplot+[orange!90!black,thick,mark=triangle*] table[x=n,y=ip,col sep=comma]{data/euclidean_corrections.csv};\addlegendentry{Interior-point}
\addplot+[violet,thick,mark=diamond*] table[x=n,y=active,col sep=comma]{data/euclidean_corrections.csv};\addlegendentry{Active-set}
\end{axis}
\end{tikzpicture}
\caption{Median Euclidean correction from the nominal command for feasible QPs.}
\label{fig:correction}
\end{figure}

The modification of the nominal command was evaluated using the Euclidean correction norm
\begin{equation}
\Delta u=\|u-u_{\mathrm{nom}}\|_2.
\label{eq:benchmark_correction}
\end{equation}
Figure~\ref{fig:correction} compares the correction magnitudes of the different algorithms. The solutions exhibit nearly identical correction magnitudes, as expected when solving the same strictly convex QP. Thus, the computational advantage of the proposed methods is achieved without intentionally increasing the modification of the nominal input.

The correction norm measures control intervention rather than safety. Feasibility is determined separately by the CBF inequality and actuator bounds. Across feasible sweep solutions, the most negative observed CBF residual was $-5.63\times10^{-12}$, consistent with floating-point rounding, and no actuator-bound violations were recorded. The reported runtimes are MATLAB measurements and do not constitute hardware worst-case execution-time guarantees. Likewise, ordinary numerical residual checks do not establish the rigorous floating-point acceptance guarantees discussed in Section~\ref{sec:verification}.

\subsection{Flight Envelope Protection Case: Angle-of-Attack Limit}

A simplified longitudinal flight-control problem is considered to demonstrate the proposed safety filter under actuator constraints. A simplified linear pitch model describes the angle-of-attack and pitch-rate dynamics:
\begin{equation}
\label{eq:flight_model}
\begin{aligned}
\dot\alpha &= q,\\
\dot q &= -d q+c^\top u,
\end{aligned}
\end{equation}
where $\alpha$ is the angle of attack, $q$ is the pitch rate, $d>0$ is the pitch-rate damping coefficient, and $c\in\mathbb R^4$ contains the control effectiveness coefficients of four independent control surfaces. Additional longitudinal states and aerodynamic coupling are neglected to focus on the actuator-constrained safety-filter problem.

A cascaded linear controller provides nominal angle-of-attack tracking. The outer loop generates a desired pitch rate from the tracking error, while the inner loop determines the desired pitch acceleration:
\begin{equation}
\label{eq:nominal_flight_controller}
\begin{aligned}
q_{\mathrm{ref}} &= K_\alpha(\alpha_{\mathrm{ref}}-\alpha),\\
\dot q_{\mathrm{des}} &= K_q(q_{\mathrm{ref}}-q).
\end{aligned}
\end{equation}
Here, $K_\alpha,K_q>0$ are controller gains. From \eqref{eq:flight_model}, the required control contribution is $\nu_{\mathrm{des}}=\dot q_{\mathrm{des}}+dq$. Weighted minimum-norm control allocation yields the nominal actuator command
\begin{equation}
\label{eq:flight_allocation}
u_{\mathrm{nom}}
=\frac{W^{-1}c}{c^\top W^{-1}c}\nu_{\mathrm{des}}.
\end{equation}

Flight-envelope protection is represented by the upper angle-of-attack constraint $h(\alpha)=\alpha_{\max}-\alpha\geq0$, with $\alpha_{\max}=0.24\,\mathrm{rad}$. Since the control input enters the second derivative of $h$, a relative-degree-two higher-order CBF (HOCBF) is employed. Defining $\psi_1=-q+k_1h$, the condition $\dot\psi_1+k_2\psi_1\geq0$ yields
\begin{equation}
\label{eq:flight_hocbf}
-c^\top u
\geq (k_1+k_2-d)q-k_1k_2h.
\end{equation}
The corresponding continuous-time HOCBF conditions require $h(0)\geq0$ and $\psi_1(0)=-q(0)+k_1h(0)\geq0$, together with satisfaction of \eqref{eq:flight_hocbf} along the trajectory.

Thus, the flight-envelope constraint has the single affine input inequality required by the proposed solution algorithms.

The simulations use $d=0.9$, $K_\alpha=1.7$, $K_q=2.8$, $k_1=2$, and $k_2=3$, with $c=[0.75,\;0.65,\;-0.55,\;-0.70]^\top$ and $W=\mathrm{diag}(1,1,1.2,1.2)$. The four surfaces are subject to deflection limits of $\pm0.50\,\mathrm{rad}$ and discrete-time command-rate limits of $\pm2.0\,\mathrm{rad/s}$. At each sampling instant, the input-constrained HOCBF-QP is solved using the proposed breakpoint-sweep algorithm (Algorithm~1), employing the incremental formulation with $T_s=0.01\,\mathrm{s}$.

The angle-of-attack reference increases from $0.10$ to $0.39\,\mathrm{rad}$ for $3\leq t<14\,\mathrm{s}$, deliberately exceeding $\alpha_{\max}$. A separate unfiltered simulation uses the same nominal controller, initial conditions, actuator constraints, and numerical integration method, but without the HOCBF constraint.

\begin{figure*}[t]
\centering
\begin{tikzpicture}
\begin{groupplot}[
    group style={
        group size=2 by 4,
        horizontal sep=1.0cm,
        vertical sep=0.75cm
    },
    width=1\columnwidth,
    height=0.13\textheight,
    xmin=0, xmax=22,
    grid=major,
    tick label style={font=\footnotesize},
    label style={font=\footnotesize},
    title style={font=\footnotesize},
    legend style={
        font=\scriptsize,
        draw=black,
        fill=white
    }
]

\nextgroupplot[
    title={(a) With proposed safety filter},
    ylabel={$\alpha$ [rad]},
    ymin=0, ymax=0.45
]
\addplot[blue, thick]
table[x=t_s, y=alpha_rad, col sep=comma]
{data/flight.csv};
\addlegendentry{Response}

\addplot[black, dotted, thick]
table[x=t_s, y=alphaRef_rad, col sep=comma]
{data/flight.csv};
\addlegendentry{$\alpha_{\mathrm{ref}}$}

\addplot[red, dashed, thick]
coordinates {(0,0.24) (22,0.24)};
\addlegendentry{$\alpha_{\max}$}

\nextgroupplot[
    title={(b) Without safety filter},
    ymin=0, ymax=0.45
]
\addplot[gray!75!black, thick]
table[x=t_s, y=alphaUnfiltered_rad, col sep=comma]
{data/flight.csv};
\addlegendentry{Response}

\addplot[black, dotted, thick]
table[x=t_s, y=alphaRef_rad, col sep=comma]
{data/flight.csv};
\addlegendentry{$\alpha_{\mathrm{ref}}$}

\addplot[red, dashed, thick]
coordinates {(0,0.24) (22,0.24)};
\addlegendentry{$\alpha_{\max}$}

\nextgroupplot[
    ylabel={$q$ [rad/s]},
    ymin=-0.3, ymax=0.5
]
\addplot[blue, thick]
table[x=t_s, y=q_rad_s, col sep=comma]
{data/flight.csv};


\nextgroupplot[
    ymin=-0.3, ymax=0.5
]
\addplot[gray!75!black, thick]
table[x=t_s, y=qUnfiltered_rad_s, col sep=comma]
{data/flight.csv};


\nextgroupplot[
    ylabel={$u$ [rad]},
    ymin=-0.6, ymax=0.6
]
\addplot[blue, thick]
table[x=t_s, y=u1, col sep=comma]
{data/flight.csv};
\addlegendentry{Surface 1}

\addplot[orange!85!black, thick]
table[x=t_s, y=u2, col sep=comma]
{data/flight.csv};
\addlegendentry{Surface 2}

\addplot[yellow!60!black, thick]
table[x=t_s, y=u3, col sep=comma]
{data/flight.csv};
\addlegendentry{Surface 3}

\addplot[purple, thick]
table[x=t_s, y=u4, col sep=comma]
{data/flight.csv};
\addlegendentry{Surface 4}

\addplot[red, dotted, thick]
coordinates {(0,0.5) (22,0.5)};
\addplot[red, dotted, thick]
coordinates {(0,-0.5) (22,-0.5)};
\addlegendentry{Limits}

\nextgroupplot[
    ymin=-0.6, ymax=0.6
]
\addplot[blue, thick]
table[x=t_s, y=u1Unfiltered, col sep=comma]
{data/flight.csv};
\addlegendentry{Surface 1}

\addplot[orange!85!black, thick]
table[x=t_s, y=u2Unfiltered, col sep=comma]
{data/flight.csv};
\addlegendentry{Surface 2}

\addplot[yellow!60!black, thick]
table[x=t_s, y=u3Unfiltered, col sep=comma]
{data/flight.csv};
\addlegendentry{Surface 3}

\addplot[purple, thick]
table[x=t_s, y=u4Unfiltered, col sep=comma]
{data/flight.csv};
\addlegendentry{Surface 4}

\addplot[red, dotted, thick]
coordinates {(0,0.5) (22,0.5)};
\addplot[red, dotted, thick]
coordinates {(0,-0.5) (22,-0.5)};
\addlegendentry{Limits}

\nextgroupplot[
    ylabel={$h$ [rad]},
    xlabel={Time [s]},
    ymin=-0.2, ymax=0.3
]
\addplot[blue, thick, forget plot]
table[x=t_s, y=hFiltered_rad, col sep=comma]
{data/flight.csv};

\addplot[red, dashed, thick]
coordinates {(0,0) (22,0)};
\addlegendentry{Safety boundary}

\nextgroupplot[
    xlabel={Time [s]},
    ymin=-0.2, ymax=0.3
]
\addplot[gray!75!black, thick, forget plot]
table[x=t_s, y=hUnfiltered_rad, col sep=comma]
{data/flight.csv};

\addplot[red, dashed, thick]
coordinates {(0,0) (22,0)};
\addlegendentry{Safety boundary}

\end{groupplot}
\end{tikzpicture}
\caption{Illustrative angle-of-attack tracking with and without safety filtering.}
\label{fig:flight}
\end{figure*}
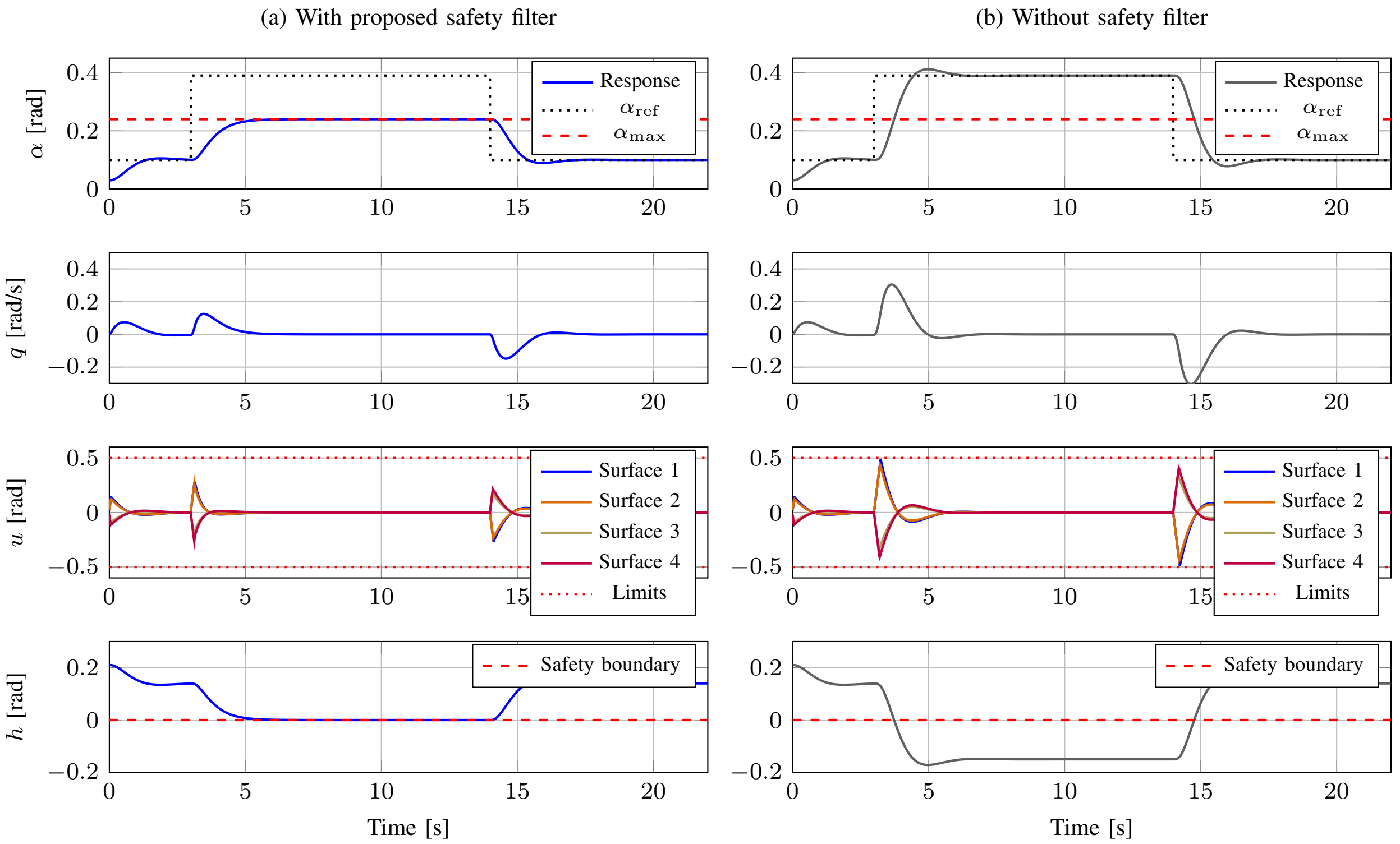

Figure~\ref{fig:flight} compares the closed-loop responses with the proposed finite-step safety filter (left) and without safety filtering (right). From top to bottom, the rows show the angle of attack, pitch rate, four control-surface deflections, and safety function $h$, respectively. Without filtering, the angle of attack reaches approximately $0.412\,\mathrm{rad}$, exceeding the prescribed limit by $0.172\,\mathrm{rad}$. In contrast, the filtered response approaches the constraint boundary without violating it at the recorded sampling instants. Accordingly, $h$ becomes negative without the filter but remains nonnegative for the filtered system.

All $2{,}200$ pointwise QPs remained feasible, and the filter modified the nominal command at $1{,}088$ sampling instants. Both simulations respected the imposed actuator magnitude and sampled rate limits, with the latter becoming active during transient responses. The computed HOCBF residual remained nonnegative up to numerical roundoff. These results illustrate the effectiveness of the proposed finite-step solution in an actuator-constrained flight-control problem, although sampled-data constraint satisfaction alone does not establish continuous-time forward invariance.

\section{Conclusion}

This paper presented finite-step algorithms for input-constrained CBF safety filters, providing exact optimal solutions with at most $2n$ breakpoint events or $n$ redistribution passes for admissible nominal inputs. The framework combines explicit feasibility assessment with verified floating-point constraint evaluation. Numerical results demonstrate computational advantages over general-purpose QP solvers and illustrate actuator-constrained flight-envelope protection. Building on established continuous quadratic knapsack methods, the approach provides predictable algorithmic effort, while closed-loop safety and hardware execution-time guarantees require separate analysis.

\bibliographystyle{IEEEtran}
\bibliography{references}
\end{document}